\documentclass[10pt,twocolumn,letterpaper]{article}

\usepackage[pagenumbers]{cvpr} % To force page numbers, e.g. for an arXiv version

\usepackage[dvipsnames]{xcolor}

\definecolor{cvprblue}{rgb}{0.21,0.49,0.74}
\usepackage{multirow}
\usepackage{amsfonts}
\usepackage{amsthm,amsmath,amssymb}
\usepackage{mathrsfs}
\usepackage[pagebackref,breaklinks,colorlinks,citecolor=cvprblue]{hyperref}

\def\paperID{11780} % *** Enter the Paper ID here
\def\confName{CVPR}
\def\confYear{2024}

\title{EFCNet: Every Feature Counts for Small Medical Object Segmentation}

\author{%
Lingjie Kong\thanks{Equal contribution.}, 
  Qiaoling Wei$^{*}$, 
  Chengming Xu, 
  Han Chen$^{ \dagger} $, 
  Yanwei Fu\thanks{Corresponding Author.} \\
  Fudan University, Shanghai, China~~~\\
}

\begin{document}
\maketitle

\begin{abstract}

This paper explores the segmentation of very small medical objects with significant clinical value. While Convolutional Neural Networks (CNNs), particularly UNet-like models, and recent Transformers have shown substantial progress in image segmentation, our empirical findings reveal their poor performance in segmenting the small medical objects and lesions concerned in this paper. This limitation may be attributed to information loss during their encoding and decoding process.
In response to this challenge, we propose a novel model named EFCNet for small object segmentation in medical images. Our model incorporates two modules: the Cross-Stage Axial Attention Module (CSAA) and the Multi-Precision Supervision Module (MPS). These modules address information loss during encoding and decoding procedures, respectively.
Specifically, CSAA integrates features from all stages of the encoder to adaptively learn suitable information needed in different decoding stages, thereby reducing information loss in the encoder. On the other hand, MPS introduces a novel multi-precision supervision mechanism to the decoder. This mechanism prioritizes attention to low-resolution features in the initial stages of the decoder, mitigating information loss caused by subsequent convolution and sampling processes and enhancing the model's global perception.
We evaluate our model on two benchmark medical image datasets. The results demonstrate that EFCNet significantly outperforms previous segmentation methods designed for both medical and normal images.

%
 % models cannot handle the information loss during the auto-encoding process. As a result, they generally perform poorly on small  objects and lesions, which are extremely common in medical research, thus making them impractical. 
 %
\end{abstract}

\section{Introduction}
\label{sec:intro}

% Small medical objects are common in disease research, such as HyperReflective Dots (HRDs) on Optical Coherence Tomography (OCT). Studies have proved that these small lesions are of great significance to medical diagnosis and treatment \cite{qin2021hyperreflective,huang2021algorithm,arthi2021hyperreflective,chung2019role}. However, manually labeling small objects in these medical images is time-consuming and labor-intensive, which is a great waste of medical resources. Therefore, there is a great need to automatically segment small medical objects with the help of AI algorithms. Fortunately, with the rapid development of deep learning methods, many Convolution neural networks (CNNs) have been proposed for medical image segmentation, such as U-Net~\cite{ronneberger2015u}, ResUNet~\cite{8589312}, DenseUNet~\cite{8697107}, ResUNet++~\cite{jha2019resunet} and MultiResUNet~\cite{ibtehaz2020multiresunet}. 

Small medical objects, like HyperReflective Dots (HRDs) observed on Optical Coherence Tomography (OCT), are frequently encountered in disease research. Various studies~\cite{qin2021hyperreflective,huang2021algorithm,arthi2021hyperreflective,chung2019role} have confirmed the significant relevance of these small lesions to medical diagnosis and treatment.
However, the manual labeling of these small objects in medical images is a time-consuming and labor-intensive task, representing a substantial drain on medical resources. Consequently, there is a pressing need to automate the segmentation of small medical objects using computer vision algorithms. Considering the type of task, one would easily recall the numerous image segmentation models such as U-Net~\cite{ronneberger2015u}, ResUNet~\cite{8589312}, DenseUNet~\cite{8697107}, ResUNet++~\cite{jha2019resunet}, TransFuse~\cite{zhang2021transfuse} and Swin-Unet~\cite{cao2022swin}, as well as the recent SAM~\cite{kirillov2023segment}. Given their generally desirable performance, it is straightforward to ask: \textit{Can these method solve the small medical objects segmentation?}

\begin{figure}
    \centering
    \includegraphics[width=8.3cm]{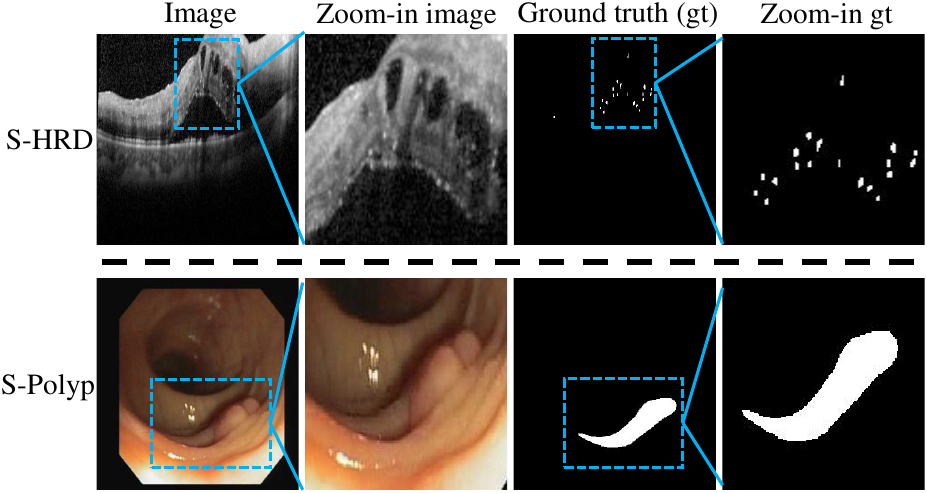}
       \vspace{-0.1in}
    \caption{Samples of small medical objects in two datasets (S-HRD and S-Polyp) in our work. \label{v_explan} }
      \vspace{-0.15in}
\end{figure}

\begin{figure*}
    \centering
    \includegraphics[width=17cm]{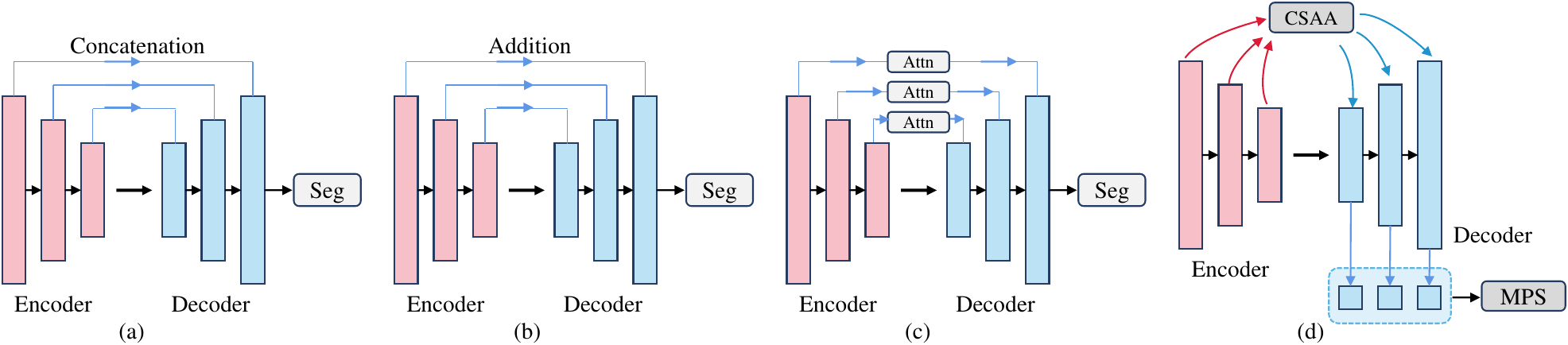}
        \vspace{-0.1in}
    \caption{Comparison of our method (d) against conventional encoder-decoder based method in previous works (a)-(c). (a)\&(b) In traditional methods~\cite{ronneberger2015u,8589312}, the single-stage features of the encoder and the corresponding single-stage features of the decoder are fused by concatenation or addition, with one segmentation head at the end of the decoder. (c) Some methods~\cite{oktay2018attention,zhang2022isnet} attempt to add attention mechanisms to the encoder, which however are limited to single-stage features. And only one segmentation head is adopted at the end of the decoder. (d) Our method aggregates all features in each stages of the encoder through CSAA to guide the decoding procedure. Multi-resolution features in each stages of the decoder are segmented with multi-precision by multiple segmentation heads through MPS.     \label{in_fig}}
    \vspace{-0.15in}
\end{figure*}

Sadly, these famous methods are generally wiped out in this specific task. Moreover, research specifically dedicated to the segmentation of small medical objects is lacking. Identified issues with previous segmentation methods reveal significant information loss, as highlighted in two points: 1) Earlier methods~\cite{ronneberger2015u,jha2020doubleu} often use features from the preceding stage in the decoder and the corresponding stage in the encoder, limiting the direct use of features for each decoding stage. This contradicts previous studies~\cite{poudel2018contextnet,cheng2020cascadepsp} indicating improved segmentation accuracy with both shallow and deep features and results in untapped information from various encoder stages. 2) Many prior approaches~\cite{8589312,jha2020doubleu,huang2020unet} employ only one segmentation head for supervision at the last decoder stage. In contrast, studies~\cite{zhang2018exfuse,chen2014semantic,chen2017deeplab} highlight the strong global perception of low-resolution features in early decoding stages, valuable for small object localization. However, information in these early decoder stages is somewhat lost during convolution and upsampling processes. Additionally, small medical objects, compared to standard-sized objects, carry less information, intensifying the impact of information loss on segmentation accuracy. Some samples of small medical objects are illustrated in Fig.~\ref{v_explan}. Notably, the SAM model~\cite{kirillov2023segment} performs inadequately in addressing this segmentation task.

% 3)Furthermore, compared with objects of normal size, small medical objects contain less information, which means that information loss of a model will have a more serious impact on the segmentation accuracy. And this is a typical segmentation task that the SAM model~\cite{kirillov2023segment} does not work well.
%
%Therefore the information loss caused by the previous methods has a great negative impact on small medical object segmentation.

% To alleviate the problem of information loss during segmentation, we propose a novel solution, which is based on the autoencoder structure. Our model pays enough attention to all the features in each stage of both encoder and decoder, thereby improving the accuracy of small medical object segmentation. Fig.~\ref{in_fig} shows the difference between our method and previous works. 

To address the challenge of information loss during segmentation, we introduce a novel solution based on the encoder-decoder structure. Our model meticulously attends to all features in each stage of both the encoder and decoder, enhancing the accuracy of small medical object segmentation. Figure~\ref{in_fig} illustrates the distinctions between our approach and prior methods.
Specifically, for the encoder, we present the Cross-Stage Axial Attention Module (CSAA), leveraging the attention mechanism to integrate features from all stages. This adaptation enables the model to dynamically learn information necessary for each decoding stage. CSAA facilitates direct reference to all valuable information in the encoder during the decoding process of each stage, mitigating information loss in the encoder.
Simultaneously, we introduce the Multi-Precision Supervision Module (MPS) for the decoder. This module adds segmentation heads with varying precision for supervision after each stage in the decoder. Low-precision segmentation heads focus on low-resolution features, temporarily overlooking local details to leverage their robust global perception. With MPS, the model effectively exploits information from each stage in the decoder, reducing information loss in this part of the model.

% Concretely, for the encoder, we proposed the Cross-Stage Axial Attention Module (CSAA), which use the attention mechanism to integrate all the features of all stages in the encoder, so that the model can adaptively learn information needed for each decoding stage. Through CSAA, the decoding process of each stage in the decoder can directly refer to all useful information in the encoder, thereby alleviating the information loss of the encoder. Meanwhile, we propose the Multi-Precision Supervision Module (MPS) for the decoder, which adds segmentation heads with different precision for supervision after each stage in the decoder. Low-precision segmentation heads are utilized for low-resolution features in order to temporarily ignore the pursuit of local details and give full play to the advantages of its strong global perception. With the help of MPS, the model can sufficiently exploit information contained in the features of each stage in the decoder, thereby alleviating the information loss in the decoder. 

To validate the efficacy of the proposed method, we carry out comprehensive experiments on two datasets S-HRD and S-polyp, as illustrated in Fig.~\ref{v_explan}. These datasets comprise a fundus Optical Coherence Tomography (OCT) image dataset created by our team and a subset from CVC-ClinicDB~\cite{bernal2015wm}. Experimental results on these two datasets demonstrate that our method outperforms previous state-of-the-art models in terms of dice similarity coefficient (DSC) and intersection over union (IoU).

% Our contributions can be summarized as follows: \\
% 1. We propose a novel idea to face the challenge of small medical object segmentation. We believe that every feature counts for small object segmentation in medical images. Our model pays attention to all features of each stage in the model and mines various information contained in them, thereby reducing information loss of small medical objects. \\
% 2. We design two modules named Cross-Stage Axial Attention Module (CSAA) and Multi-Precision Supervision Module (MPS), to effectively reduce the information loss in the encoder and decoder respectively and therefore improve segmentation accuracy of the model. \\
% 3. We construct a new benchmark for the study of small medical objects segmentation. And our experiments on two datasets of small medical objects demonstrate that our model outperforms the previous state-of-the-art models to a large extent.

Our contributions are outlined as follows: \\
1. \textbf{Innovative Segmentation Approach}: We introduce a novel concept to address the challenge of small medical object segmentation. Emphasizing the significance of every feature in medical images, our model meticulously attends to all features at each stage. This approach enables the extraction of diverse information, thereby mitigating information loss associated with small medical objects. \\
2. \textbf{Proposed Modules for Enhanced Accuracy}: We devise two key modules, namely the Cross-Stage Axial Attention Module (CSAA) and the Multi-Precision Supervision Module (MPS). These modules effectively tackle information loss in the encoder and decoder, respectively, resulting in an improved segmentation accuracy for the model. \\
3. \textbf{Benchmark Construction and Model Validation}: We establish a new benchmark for evaluating small medical object segmentation. Through experiments on two datasets focusing on small medical objects, our model significantly outperforms previous state-of-the-art models. This demonstrates the robustness and superiority of our proposed approach in this challenging domain.

\section{Related Works}
\noindent \textbf{Medical Image Segmentation}.
Recently, researchers have introduced several innovative methods~\cite{chen2018drinet,milletari2016v,gu2019net,hatamizadeh2022unetr,valanarasu2022unext} for semantic segmentation in medical images. Zhang \etal~\cite{zhang2021transfuse} 
proposed a unique hybrid structure that concurrently integrates CNN and Transformer, leading to a reduction in the loss of low-level details.
%
%proposed a novel hybrid structure that combines CNN and Transformer in parallel, which %reduced the loss of low-level details. 
%
Chen \etal~\cite{chen2021transunet} incorporated a Transformer module into the U-Net encoder, enhancing the model's ability for long-range modeling.
%
%embedded Transformer module into the encoder of U-Net, which improved the long-range %modeling ability of the model. 
%
Wang \etal~\cite{wang2022stepwise} addressed overfitting by employing a Transformer encoder and introduced the progressive locality decoder to improve local information processing in medical images.
%
%used a Transformer encoder to alleviated the overfitting phenomenon, and designed %Progressive Locality Decoder to enhance local information of medical images. 
%
%Although these methods above have played a great role in medical image segmentation, 
%they have not fully considered the impact of object size on segmentation results. 
%
While these methods have significantly contributed to medical image segmentation, they often fall short in accounting for the impact of object size on segmentation results.
Particularly, these models tend to underperform when confronted with the segmentation of small objects. Lou \etal~\cite{lou2022caranet} recognized the significance of considering the size of medical objects in the segmentation process, and 
introduced a Context Axial Reverse Attention module (CaraNet) to assist the model in detecting local information related to small medical objects. However, 
the use of bilinear interpolation in the decode stage of CaraNet leads to substantial information loss, significantly affecting the segmentation of small medical objects.
%
%bilinear interpolation used in the decode stage of CaraNet will cause serious %information loss, which has a great negative impact on small medical object segmentation.
%
In contrast to the aforementioned methods, our model addresses the issue of information loss in small medical objects through CSAA and MPS, effectively improving segmentation accuracy.

% Different from methods mentioned above, our model reduces the information loss of small medical objects through CSAA and MPS, and thus effectively improves the segmentation accuracy.

\noindent \textbf{General Segmentation Model}.
% Recently, Kirillov et al.~\cite{kirillov2023segment} proposed a Segment Anything Model (SAM), a general segmentation model which has made important progress in natural image segmentation. However, due to the fine structures and complex boundaries in medical images, SAM is not suitable for most medical image segmentation tasks, including small medical object segmentation, without manual guidance \cite{huang2023segment,he2023accuracy}. Ma et al.~\cite{ma2023segment} proposed MedSAM based on SAM specifically for medical image segmentation, which has a great improvement in medical image segmentation tasks compared with SAM~\cite{kirillov2023segment}. However, MedSAM~\cite{ma2023segment} still performs poorly when faced with the task of small medical object segmentation.
%
%
Recently, Kirillov \etal~\cite{kirillov2023segment} introduced the Segment Anything Model (SAM), a versatile segmentation model that has made significant strides in the realm of natural image segmentation. Despite its success in general applications, SAM proves unsuitable for many medical image segmentation tasks due to the intricate structures and complex boundaries present, particularly in cases involving small medical objects, without manual guidance~\cite{huang2023segment,he2023accuracy}.
    In response to these limitations, Ma \etal~\cite{ma2023segment} devised MedSAM as an adaptation of SAM tailored specifically for medical image segmentation. MedSAM exhibits notable advancements in handling medical image segmentation tasks compared to SAM~\cite{kirillov2023segment}. Nevertheless, even with these improvements, MedSAM~\cite{ma2023segment} still struggles when tasked with the segmentation of small medical objects concerned in this paper.

% \noindent \textbf{Attention Mechanism}.
% Plenty of Attention-based methods \cite{oktay2018attention,shen2018reinforced,huang2019attention,sinha2020multi,tao2020hierarchical} have been proposed in recent years for various tasks in computer vision and natural language processing. Vaswani et al. \cite{vaswani2017attention} abandoned the traditional convolutional structure and  constructed a novel architecture named Transformer using attention mechanisms. Woo et al. \cite{woo2018cbam} added both Channel Attention and Spatial Attention in convolutional neural networks. Zhang et al. \cite{zhang2022epsanet} proposed the Pyramid Squeeze Attention Module (PSA) to help the model explore spatial information at different scales. Inspired by the attention-baseds method mentioned above, we propose Cross-Stage Axial Attention Module (CSAA), which facilitates the feature fusion and reduces information loss in the model.

\noindent \textbf{Attention Mechanism}.
Numerous attention-based methods~\cite{oktay2018attention,shen2018reinforced,huang2019attention,sinha2020multi,tao2020hierarchical} have emerged in recent years, applied to diverse tasks in computer vision and natural language processing. Vaswani \etal~\cite{vaswani2017attention} broke away from the conventional convolutional structure and introduced the Transformer, a novel architecture utilizing attention mechanisms. Woo \etal~\cite{woo2018cbam} enhanced convolutional neural networks by incorporating both Channel Attention and Spatial Attention. Zhang \etal~\cite{zhang2022epsanet} proposed the Pyramid Squeeze Attention Module (PSA) to enable the model to capture spatial information across different channels.
Building upon the insights gained from the aforementioned attention-based methods, we introduce the Cross-Stage Axial Attention Module (CSAA). This module facilitates feature fusion and minimizes information loss within the model.

\section{Method}
% In this chapter, we first introduce the task of small medical object segmentation and some notations in Section~\ref{setup}. Next, we introduce the overall structure of EFCNet in Section~\ref{overall}. Then, we show the details of Cross-Stage Axial Attention Module (CSAA) and Multi-Precision Supervision Module (MPS) in Section~\ref{CSAA} and Section~\ref{MPS}, respectively. Finally, we introduce the Loss Function of our model in Section~\ref{loss}.
%
%
% We begin by introducing the task of small medical object segmentation and related notations. Then, we present the overall structure of EFCNet in Sec.~\ref{overall} and explain the specifics of the Cross-Stage Axial Attention Module (CSAA) and the Multi-Precision Supervision Module (MPS) in Sec.~\ref{CSAA} and Sec.~\ref{MPS} respectively. Finally, we discuss the Loss Function of our model in Sec.~\ref{loss}.

We introduce small medical object segmentation and related notations. In Sec.\ref{overall}, we outline the overall structure of EFCNet. We detail the CSAA in Sec.~\ref{CSAA} and the MPS in Sec.~\ref{MPS}. Lastly, we discuss our loss function in Sec.\ref{loss}.

%\paragraph{Problem setup and notations}
\noindent \textbf{Problem Setup and Notations}.
%\label{setup}
In our segmentation task, we denote the medical picture dataset as $X = \{x_{1}, ..., x_{m} | x_{i}\in\mathbb{R}^{C\times H\times W}, i=1,2,..,m \} $. 
Doctors meticulously and manually annotate lesions in each image  \(x_{i}\), forming the ground truth set  $Y = \{y_{1}, ..., y_{m} | y_{i} \in \{0,1\}^{1\times H\times W}, i=1,2,..,m \} $.
%
% Lesions in each image \(x_{i}\) are carefully and manually annotated by doctors, constituting the ground truth set $Y = \{y_{1}, ..., y_{n} | y_{i} \in \{0,1\}^{1\times H\times W} \} $. 
%
The complete dataset is denoted as \(D=\{X, Y\}\), which is splitted into a training set $ D_\text{train}=\{X_\text{train}, Y_\text{train}\}$ and a testing set $D_\text{test} = \{X_\text{test}, Y_\text{test}\}$. 

Our objective is to develop an algorithm that empowers our model to effectively segment small medical objects from \(D_\text{train}\) and demonstrate robust performance on \(D_\text{test}\).

% our model can gain the ability to segment small medical objects from \(D_{train}\) and perform well in \(D_{test}\).

\subsection{Overall Architecture}
\label{overall}

\begin{figure*}[t]
    \centering
    \includegraphics[width=17.4cm]{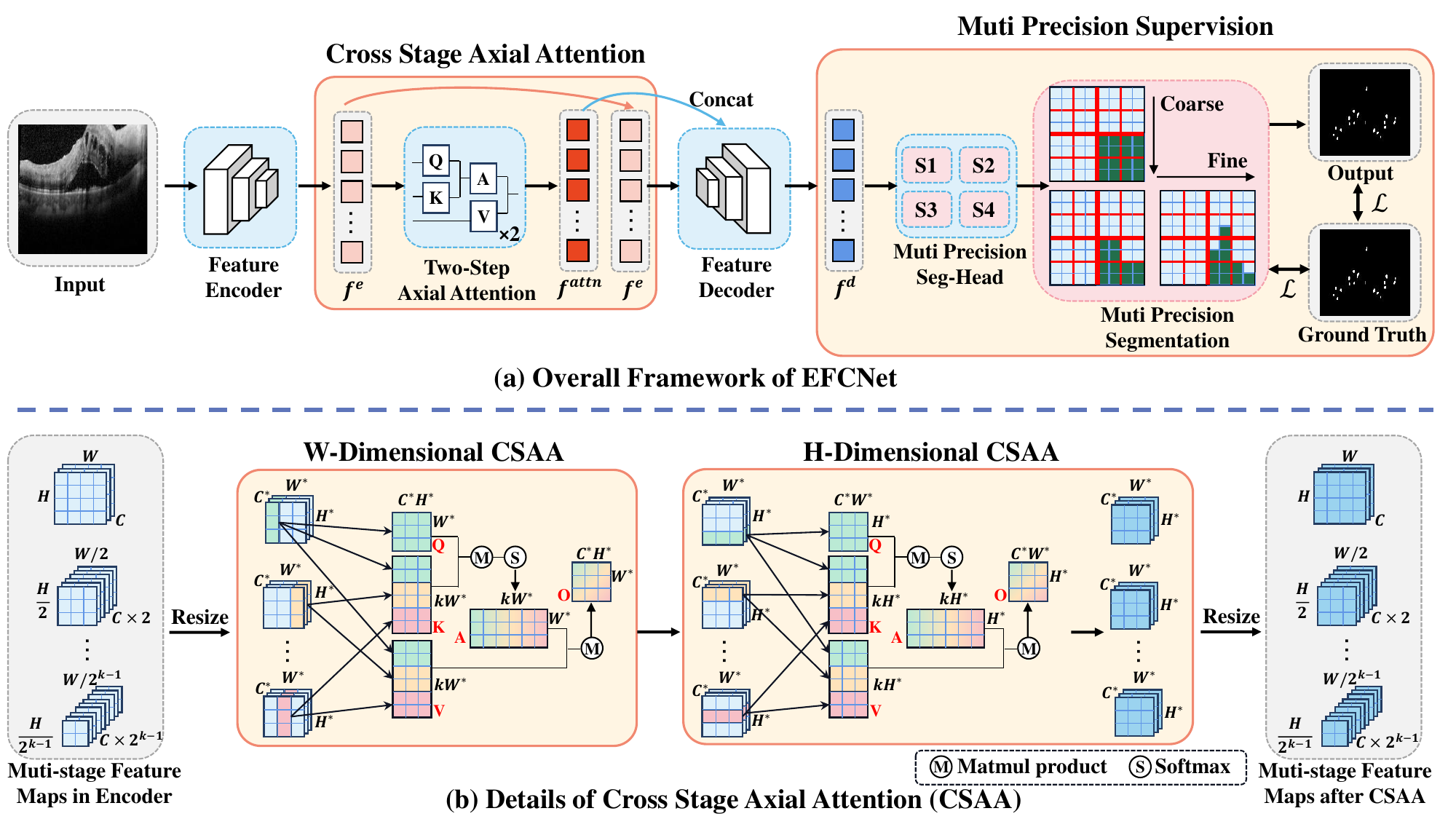}
        \vspace{-0.1in}
    \caption{Structure of our proposed EFCNet. (a) Overview of our method, featuring an Encoder-Decoder architecture equipped with the Cross-Stage Axial Attention Module (CSAA) and Multi-Precision Supervision Module (MPS). (b) Details of our Cross-Stage Axial Attention Module (CSAA). The CSAA combines features from each stage of the encoder, dynamically extracts information about small medical objects, and directs the decoding process of each stage in the decoder.  \label{all_fig} }
  
    \vspace{-0.15in}
\end{figure*}

We propose a novel method called EFCNet to address the challenge of segmenting small medical objects. Specifically, we have designed two modules, the Cross-Stage Axial Attention Module (CSAA) and the Multi-Precision Supervision Module (MPS), to ensure that the model  focuses on  small medical object infomration  in both the encoder and decoder. 

In Fig.~\ref{all_fig}(a), an image 
containing small medical objects serves as input to our model.
Initially,  the image undergoes encoding through \(k\) stages, producing feature maps encompassing diverse information about small medical objects.
%
%First, the image passes through the encoder with \(k\) stages, obtaining feature maps %containing various types of information of small medical objects. 
%
% Next, features in all stages of the encoder are put into CSAA Module, forcing the model to adaptively learn information about small medical objects that is conducive to the decoding procedure. Then, the decoder processes the feature maps stage by stage under the guidance of CSAA Module. 
%
Subsequently, the CSAA Module processes features from all encoder stages, compelling the model to adaptively learn pertinent information for the decoding phase. The decoder then sequentially processes these feature maps under the guidance of the CSAA Module. 
Lastly, the feature maps from all decoder stages enter the MPS Module to yield multi-precision prediction results, receiving separate supervisions. During testing, we utilize the segmentation output from the last decoder stage as the final result of our model.

% Finally, feature maps of all stages in the decoder are put into MPS Module to obtain multi-precision prediction results and get supervisions separately. In the testing phase, we use the segmentation output of the last stage of decoder as the final segmentation result of our model.

\subsection{CSAA Module}
\label{CSAA}

% Features in different stages of the encoder contain different kinds of information about small medical objects. However, most of the information cannot be directly used by the decoder, and some information is even lost in the convolution and downsampling process. In order to reduce the information loss in the encoder and take full advantages of the information about small medical objects, we design Cross-Stage Axial Attention Module (CSAA), which adaptively learns information from features in all stages of the encoder, and then guides the subsequent decoding work. As shown in Fig.~\ref{all_fig}, CSAA includes three steps: resizing, two-stepaxial-attention and compressing. 

Information about small medical objects is dispersed across different encoder stages, each containing varied types of data. However, much of this information is not directly usable by the decoder, and some is lost during the convolution and downsampling processes. To reduce information loss in the encoder and fully leverage insights about small medical objects, we present a novel Cross-Stage Axial Attention Module (CSAA). This module adaptively learns from features in all encoder stages and subsequently guides the decoding process. As depicted in Fig.~\ref{all_fig}(b), CSAA have four steps: resizing, W-dimensional axial attention, H-dimensional axial attention and resizing back.

\noindent \textbf{Resizing}.
To enhance the fusion of features from all encoder stages, we resize all feature maps in each encoder stage to \((C^{*},H^{*},W^{*})\), adjusting both spatial and channel dimensions through convolution operations:
%
%\begin{equation}
%    f_{i}^{*} = \sigma (BN(conv_{j}^{e}(f_{i}^{e}))) \in \mathbb R^{C^{*} \times H^{*} \times W^{*}}, i=1,2,..,k,  
%\end{equation}
%
\begin{equation}
    \begin{split}
        &f_{i}^{*} = \sigma (BN(conv_{i}^{e}(f_{i}^{e}))), i=1,2,..,k,
        %\\& \in \mathbb R^{C^{*} \times H^{*} \times W^{*}}, i=1,2,..,k,
    \end{split}  
\end{equation}
where \(f_{i}^{e}\) represents the feature map in stage \(i\) of the encoder; \(\sigma\) and \(BN\) denote ReLU and Batch Normalization respectively; and \(k\) is the number of stages in the encoder and decoder. We consider the k-stage resized feature maps \(\{f_{i}^{*}\}_{i=1}^{k}\) as the input  for the subsequent W-dimensional axial attention step.

\noindent \textbf{W-Dimensional CSAA}.
We firstly generate Query \(\{Q_{i,w}\}_{i=1}^{k}\), Key \(\{K_{i,w}\}_{i=1}^{k}\), Value \(\{V_{i,w}\}_{i=1}^{k}\) based on the feature maps \(\{f_{i}^{*}\}_{i=1}^{k}\) in the width (W) dimension:
\begin{equation}
    \label{attn_eq}
    \begin{split}
        &Q_{i,w} = W_{i}^{Q}(f_{i,w}^{*}),i=1,2,..,k, \\& K_{i,w} = W_{i}^{K}(f_{1,w}^{*},f_{2,w}^{*},...,f_{k,w}^{*}),i=1,2,..,k,\\& V_{i,w} = W_{i}^{V}(f_{1,w}^{*},f_{2,w}^{*},...,f_{k,w}^{*}),i=1,2,..,k,
    \end{split}  
\end{equation}
where \(\{W_{i}^{Q}\}_{i=1}^{k}\), \(\{W_{i}^{K}\}_{i=1}^{k}\), \(\{W_{i}^{V}\}_{i=1}^{k}\) represent the weight matrix used to generate \(\{Q_{i,w}\}_{i=1}^{k}\), \(\{K_{i,w}\}_{i=1}^{k}\), \(\{V_{i,w}\}_{i=1}^{k}\) respectively; \(k\) is the number of stages in the encoder and decoder; and \(f_{i,w}^{*}\) denotes the feature \(f_{i}^{*}\) in width dimension. Equation~\ref{attn_eq} shows that \(K_{i,w}\) and \(V_{i,w}\) merge the information from all stages of the encoder in width dimension. Next, we get the output \(\{f_{i}^{w}\}_{i=1}^{k}\) of W-Dimensional Axial-Attention by
\begin{equation}
    \begin{split}
        f_{i}^{w}=\text{Softmax}(\frac{Q_{i,w}K_{i,w}^{T}}{\sqrt{C^{*}H^{*}}})V_{i,w} ,i=1,2,..,k.
    \end{split}  
\end{equation}

\noindent \textbf{H-Dimensional CSAA}.
Similarly above,
we firstly generate Query \(\{Q_{i,h}\}_{i=1}^{k}\), Key \(\{K_{i,h}\}_{i=1}^{k}\), Value \(\{V_{i,h}\}_{i=1}^{k}\) based on the feature maps \(\{f_{i}^{w}\}_{i=1}^{k}\) in the height (H) dimension:
\begin{equation}
    \label{attnh_eq}
    \begin{split}
        &Q_{i,h} = W_{i}^{Q}(f_{i,h}^{w}),i=1,2,..,k, \\& K_{i,h} = W_{i}^{K}(f_{1,h}^{w},f_{2,h}^{w},...,f_{k,h}^{w}),i=1,2,..,k,\\& V_{i,h} = W_{i}^{V}(f_{1,h}^{w},f_{2,h}^{w},...,f_{k,h}^{w}),i=1,2,..,k,
    \end{split}  
\end{equation}
where \(\{W_{i}^{Q}\}_{i=1}^{k}\), \(\{W_{i}^{K}\}_{i=1}^{k}\), \(\{W_{i}^{V}\}_{i=1}^{k}\) represent the weight matrix used to generate \(\{Q_{i,h}\}_{i=1}^{k}\), \(\{K_{i,h}\}_{i=1}^{k}\), \(\{V_{i,h}\}_{i=1}^{k}\) respectively; \(k\) is the number of stages in the encoder and decoder; and \(f_{i,h}^{w}\) denotes the feature \(f_{i}^{w}\) in height dimension. Through Eq.~\ref{attn_eq} and Eq.~\ref{attnh_eq},we merge the information from all stages of the encoder in both width dimension and height dimension. Next, we obtain the output \(\{f_{i}^{h}\}_{i=1}^{k}\) of H-Dimensional Axial-Attention through the attention operation:
\begin{equation}
    \begin{split}
        f_{i}^{h}=\text{Softmax}(\frac{Q_{i,h}K_{i,h}^{T}}{\sqrt{C^{*}W^{*}}})V_{i,h} ,i=1,2,..,k.
    \end{split}  
\end{equation}

\noindent \textbf{Resizing back}. To aid the guidance of the decoding process with the information acquired through axial attention, we resize the output feature \(f_{i}^{h}\) of the two-step axial attention to match the dimensions of the feature map in the corresponding decoding stage \(i\), adjusting both spatial and channel dimensions through convolution operations, denoted as  \(\{f_{i}^{attn}\in\mathbb{R}^{C_{i}\times H_{i}\times W_{i}}\}_{i=1}^k\),
\begin{equation}
    \begin{split}
        &f_{i}^{attn} = \sigma (BN(conv_{i}^{h}(f_{i}^{h}))), i=1,..,k,
    \end{split}  
\end{equation}
where \(\sigma\) and \(BN\) denote ReLU and Batch Normalization respectively; and \(k\) is the number of stages in the encoder and decoder. Finally, we concatenate \(f_{i}^{attn}\) to the feature map of the corresponding stage in the decoder along the channel dimension.

It is noteworthy that traditional two-dimensional attention mechanism demands substantial computing resources~\cite{wang2020axial}. To address this, we employ two-stage one-dimensional attention modules in CSAA, conducting attention processing sequentially on the feature maps in the width and height dimensions.
%
% we adopt a two-stage one-dimensional self-attention model here, performing self-attention processing on the feature maps in the Height and Width dimensions in turn.

% As can be seen, CSAA helps the model mine the  information about small medical objects in the encoder to a great extent, and assigns it to the corresponding stage of the decoder.Different from previous models, our decoding procedure of each stage in the decoder is guided by segmentation related information exploited from all stages of the encoder, with the help of CSAA. Through CSAA, our model realizes feature fusion in encoder, strengthens the connection between encoder and decoder, and reduces the loss of information in encoder to a large extent.

The CSAA significantly aids the model in extracting information about small medical objects from the encoder and appropriately allocates it to the corresponding stage of the decoder. Unlike prior models, our decoding process for each stage in the decoder is influenced by segmentation-related information gleaned from all stages of the encoder, facilitated by CSAA. Through CSAA, our model achieves feature fusion in the encoder, and reinforces the linkage between the encoder and decoder. % and substantially mitigates information loss in the encoder.

%We take stage2 as an example to show the working details of CSAA, as shown in Figure~\ref{csaa_fig}. First, we adjust the size of all feature maps in four stages of the encoder to the same size as that of stage2, i.e. \( (2\times C, \frac{H}{2}, \frac{W}{2}) \), through convolution and deconvolution operations.

%Next, we concatenate all resized features of the encoder in channel dimension to obtain a new feature \(feature_{all}\) with a size of \( (8\times C, \frac{H}{2}, \frac{W}{2}) \). Then, we put \(feature_{all}\) into an axial attention model to comprehensively consider all feature maps of each channel, and thus adaptively obtain corresponding useful information to guide the decoding process of each stage in the decoder. It is worth mentioning that considering that conventional two-dimensional self-attention consumes a lot of computing resources, we adopt a two-stage one-dimensional self-attention model here, performing self-attention processing on the feature maps in the H and W dimensions, respectively. Finally, the output of axial attention model is adjusted to the same size as that of stage2, i.e. \( (2\times C, \frac{H}{2}, \frac{W}{2}) \), through simple convolution operation, and is regarded as the final output of CSAA, which is concatenated into the corresponding stage of the decoder.

\subsection{MPS Module}
\label{MPS}

% Low-resolution features in the decoder have strong global perception, which can help the model improve performance on small medical object segmentation. In previous models ~\cite{ronneberger2015u,chen2021transunet,lou2022caranet}, however, the globally perceptual information is not fully exploited, and much of the useful information is lost in subsequent convolution and upsampling process. In order to solve this problem, we design Multi-Precision Supervision Module (MPS) to mine the information of low-resolution features in the decoder and reduce its loss in subsequent decoding process. As shown in Fig.~\ref{all_fig}, MPS includes two steps: segmentation and upsampling.

Low-resolution features in the decoder possess robust global perception, enhancing the model's performance in small medical object segmentation. However, in prior models such as~\cite{ronneberger2015u,chen2021transunet,lou2022caranet}, the globally perceptual information is not fully harnessed; and a significant amount of useful information is lost in the subsequent convolution and upsampling processes. To tackle this issue, we introduce the Multi-Precision Supervision Module (MPS) to extract information from low-resolution features in the decoder and diminish its loss in the ensuing decoding process.  Specifically, MPS consists of two steps: segmentation and upsampling.

%Illustrated in Fig.\ref{all_fig},

\noindent \textbf{Segmentation}.
To thoroughly extract information about small medical objects from each stage of the decoder, we individually feed feature maps from each decoder stage into corresponding segmentation heads. This process yields segmentation results with distinct resolutions, denoted as  \(\{P_{i}\in\mathbb{R}^{C_{i}\times H_{i}\times W_{i}}\}_{i=1}^k\),
% In order to mine the information of small medical object contained in each stage of the decoder adequately, we sent feature maps in each stage of the decoder to corresponding segmentation heads separately, and thus we get segmentation results with different resolutions
\begin{equation}
    \begin{split}
        &P_{i} = S(\sigma (BN(conv_{i}^{d}(f_{i}^{d})))),\quad i=1,..,k, 
    \end{split}
\end{equation}
where \(f_{i}^{d}\) represents the feature map in stage \(i\) of the decoder; \(\sigma\) and \(BN\) denote ReLU and Batch Normalization respectively; and \(S(.)\) represents the sigmoid function.

\noindent \textbf{Upsampling}.
We employ  neighbor interpolation method to upsample the segmentation results obtained in the previous step to match the size same of the ground truth image. This process enables us to achieve multi-precision segmentation \(\{M_{i}\}_{i=1}^k\) for small medical objects, as illustrated in Fig.~\ref{all_fig}(a).
\begin{equation}
   M_{i} = Upsample(P_{i})\in \mathbb R^{C \times H \times W}, i=1,2,..,k.
\end{equation}
We oversee the segmentation results of different precision with the ground truth label, ensuring that each stage of the decoder encompasses sufficient information to facilitate the segmentation of small medical objects.

% As mentioned earlier, we formulate 
% a supervision  with different precision for features of different resolutions. Considering that low-resolution features have strong global perception but lack local details, we use low-precision supervision for them in order to temporarily give up the pursuit of local details but give full play to the advantages of low-resolution features with strong global perception. This multi-precision supervision approach preserves the advantages of the traditional single-segmentation head while retaining more global perception from low-resolution features for the model, thereby improving the performance on small medical objects segmentation.

In MPS, we formulate a supervision strategy with varying precision for features of different resolutions. Recognizing that low-resolution features possess robust global perception but lack local details, we employ low-precision supervision for them. This decision is made to temporarily forego the emphasis on local details while capitalizing on the strengths of low-resolution features with powerful global perception. This multi-precision supervision approach preserves the advantages of the conventional single-segmentation head while preserving additional global perception from low-resolution features. Consequently, it enhances the model's performance in small medical object segmentation.

\subsection{Loss function \label{loss}}

Considering that the positive and negative pixels are extremely unbalanced in small medical object segmentation tasks, we adopt a combination of DiceLoss~\cite{sudre2017generalised} and Binary Cross Entropy(BCE) Loss during the training process. The loss for the segmentation maps produced by each stage of the decoder is set as follows.
%For the segmentation maps output by each stage of the decoder, we set the loss as:
\begin{equation}
    \begin{split}
        &\mathcal{L}_{i} = \lambda_{1}\cdot \mathcal{L}_\text{Dice} (M_{i}, Y) + \lambda_{2}\cdot \mathcal{L}_\text{BCE} (M_{i}, Y),
    \end{split}
\end{equation}
where $i=1,2,..,k$ denotes stage indexes; \(M_{i}\) represents the result predicted by the model at stage \(i\); and \(Y\) represents the ground truth with the hyperparameters \(\lambda_{1}\) and \(\lambda_{2}\)  to balance DiceLoss and BCELoss.
%We set \(\lambda_{1}=0.7\) and \(\lambda_{2}=0.3\) in the experiments.
Taking into account all segmentation results output by each stage of the decoder, the total loss of the model is
\begin{equation}
    \mathcal{L}_\text{total} = \sum_{i=1}^k\alpha_{i}\cdot \mathcal{L}_{i}, 
\end{equation}
where  hyperparameter \(\{\alpha _{i}\}_{i=1}^{k}\) leverage the losses of segmentation results with different precision.
%We set \(alpha_{1}\), \(alpha_{2}\), \(alpha_{3}\), \(alpha_{4}\) to 0.7, 0.8, 0.9, 1.0 respectively in the experiments.

\section{Experiment}

%\noindent \textbf{Experimental Setup}.
% \noindent \textbf{S-HRD Dataset}.
% We collect 313 optical coherence tomography (OCT) images of macular edema patients and aim to segment the small HyperReflective Dots (small HRDs) in them. The dataset is dubbed as S-HRD, and S is for 'Small'. In S-HRD, the area of each lesion is less than 1 percent of the size of an entire image. We confirm that appropriate consent has been obtained for the use and display of images in our research. And we are fully prepared to make the dataset publicly available. More details of data collection process and privacy concerns are introduced in Sec. xxx? of Supplementary Material.

\noindent \textbf{S-HRD Dataset}.
            We have gathered a dataset comprising 313 optical coherence tomography (OCT) images from patients with macular edema, with the objective of segmenting small HyperReflective Dots (small HRDs) within them. We refer to this dataset as S-HRD, where 'S' indicates 'Small'. In S-HRD, the area of each lesion is less than 1 percent of the entire image size. All the ground truths have been manually labeled by experienced eye doctors with over ten years of expertise. We have ensured that appropriate consent has been obtained for the utilization and presentation of images in our research. For further insights into the data collection process, privacy considerations and relevant medical knowledge, please refer to the Supplementary.

% \noindent \textbf{S-Polyp Dataset}.
% We select a polyp segmentation dataset CVC-ClinicDB \cite{bernal2015wm}, delete images containing large medical lesions, and keep 229 images in which all lesions are small medical objects. We name this dataset S-Polyp. The area of each lesion is less than 5 percent of the size of an entire image. Some examples of S-HRD and S-Polyp are shown in Fig.~\ref{v_explan}.

\noindent \textbf{S-Polyp Dataset}.
We build a small-polyp segmentation dataset by excluding images with sizable medical lesions in CVC-ClinicDB~\cite{bernal2015wm}. From this selection, we retain 229 images where all lesions are small medical objects. We label this dataset as S-Polyp. In S-Polyp, the area of each lesion is less than 5 percent of the entire image size. Examples of both S-HRD and S-Polyp are illustrated in Fig.~\ref{v_explan}.

To mitigate limitations and address the specificity of the two datasets, we employ a five-fold cross-validation approach to assess the performance of our model.
% In order to avoid limitation and particularity of the two datasets, we use five-fold cross-validation to evaluate the performance of our model. 

% \noindent \textbf{Definition of Small Medical Objects}. Since there is no uniform definition of small medical objects in previous works, we define it in our work: If the ratio of the number of pixels \(n\) in an object to the number of pixels \(N\) in the entire picture is less than 5\(\%\), then the object is called a small medical object. If the ratio is less than 1\(\%\), we call the object an extremely small medical object. In S-Polyp, all objects are small medical objects. In S-HRD, all objects are extremely small medical object.

\noindent \textbf{Definition of Small Medical Objects}.
Given the absence of a consistent definition for small medical objects in previous works, we establish our own criteria. 
 In our framework, an object is considered a small medical object if the ratio of its pixel count 
\(n\) to the total pixel count 
\(N\)  in the entire image is less than 5\(\%\). For objects with a ratio below 1\(\%\), we classify them as extremely small medical objects. In the S-Polyp dataset, all objects fall under the category of small medical objects, while in S-HRD, all objects are classified as extremely small medical objects.

\noindent \textbf{Evaluation Metrics}.
We use two common metrics to compare our model with previous state-of-the-art models.
Dice Similariy Coefficient (DSC) is defined as:
\begin{equation}
    DSC =\frac{2 \times |P\cap G|}{|P|+|G|}, 
\end{equation}
where \(P\) represents the area of the predicted label and \(G\) represents the  area of the ground truth.
Intersection over Union (IoU) is defined as:
\begin{equation}
    IoU =\frac{S_{i}}{S_{u}}, 
\end{equation}
where \(S_{i}\) represents the area where the predicted label and  ground truth overlap; and \(S_{u}\) for the total area of the two.

\noindent \textbf{Implementation Details}.
% Our experiments are implemented on an NVIDIA RTX A6000 GPU with 48GB of memory. We use the SGD optimizer with an initial learning rate of 0.01. Our training process lasts for 200 epochs with a batch size of 4. All input images are uniformly resized to 352\(\times\)352. The number of stages in encoder and decoder of our model is set to 4. 
%
We conduct our experiments using one NVIDIA RTX A6000 GPU equipped with 48GB of memory. The SGD optimizer is employed with an initial learning rate of 0.01. Our training spans 200 epochs, employing a batch size of 4. All input images are resized uniformly to $352\times352$. The model configuration includes 4 stages in both the encoder and decoder. We set \(\lambda_{1}\) and \(\lambda_{2}\) to 0.7 and 0.3 respectively to  balance DiceLoss and BCELoss. And \(\alpha_{1}\), \(\alpha_{2}\), \(\alpha_{3}\), \(\alpha_{4}\) are set to 1.0, 0.9, 0.8,0.7  respectively to balance losses of muti-precision segmentation results.

% Since there are few works specially dedicated to small medical object segmentation, we choose recent models in the field of medical image segmentation as a reference, including state-of-the-art models: 
\noindent \textbf{Competitors}.
Given the limited number of works dedicated specifically to small medical object segmentation, we reference recent models from the broader field of medical image segmentation, including state-of-the-art models:
(1) CNN based methods: U-Net~\cite{ronneberger2015u}, Attention-UNet~\cite{oktay2018attention}, MSU-Net~\cite{su2021msu}, CaraNet~\cite{lou2022caranet}; (2) Transformer based methods: TransFuse~\cite{zhang2021transfuse}, TransUNet~\cite{chen2021transunet}, SSFormer~\cite{wang2022stepwise}, Swin-UNet~\cite{cao2022swin}; (3) Segment Anything Model (SAM)~\cite{kirillov2023segment} and the related works: SAM without any prompt, SAM with point, SAM with box and MedSAM~\cite{ma2023segment}. (4) 
Additionally, we assess the performance of an enlarged version of U-Net (U-Net-Large), where both the encoder and decoder are scaled up from 4 layers to 12 layers. This exploration aims to understand the impact of model size on segmentation accuracy.

% We also test the performance of the enlarged version of U-Net (U-Net-Large), whose encoder and decoder are simply scaled up from 4 layers to 12 layers, in order to explore the influence of model size on segmentation accuracy.

\subsection{Quantitative Results}

\begin{table*} \small
    \caption{Comparison of our EFCNet with competitors on S-HRD and S-Polyp in DSC ($\%$) and IoU ($\%$).     \label{big_tab} }    

    \vspace{-0.1in }
    \centering
    \tabcolsep=0.18cm
    \begin{tabular}{c|c|cccccc|cccccc}
        \hline
        \multirow{2}*{Metrics} & \multirow{2}*{Methods} & \multicolumn{6}{c}{S-HRD} & \multicolumn{6}{|c}{S-Polyp}\\
        \cline{3-14}
        && Fold0 & Fold1 & Fold2 & Fold3 & Fold4 & Mean & Fold0 & Fold1 & Fold2 & Fold3 & Fold4 & Mean\\
        \hline
        \multirow{14}*{DSC ($\%$)} 
        & U-Net~\cite{ronneberger2015u} & 42.09 & 36.51 & 37.74 & 35.90 & 41.29 & 38.71 & 75.90 & 75.47 & 81.95 & 72.06 & 76.73 & 76.42\\
        & U-Net-Large~\cite{ronneberger2015u} & 43.02 & 36.72 & 38.24 & 37.82 & 41.60 & 39.48 & 78.93 & 78.32 & 82.76 & 73.70 & 77.84 & 78.31\\
        & Attn-UNet~\cite{oktay2018attention} & 44.33 & 39.97 & 39.07 & 38.46 & 39.90 & 40.35 & 79.89 & 81.17 & 81.19 & 77.01 & 74.68 & 78.79\\
        & MSU-Net~\cite{su2021msu} & 44.06 & 40.94 & 38.66 & 38.54 & 39.44 & 40.33 & 87.60 & 86.64 & 86.03 & 82.99 & 75.56 & 83.76\\
        & CaraNet~\cite{lou2022caranet} & 17.03 & 13.79 & 12.46 & 13.40 & 13.07 & 13.95 & 85.12 & 78.66 & 88.14 & 75.02 & 82.56 & 81.90\\
        & TransUNet~\cite{chen2021transunet} & 32.84 & 30.34 & 30.71 & 31.54 & 32.81 & 31.65 & 84.83 & 82.13 & 88.24 & 76.53 & 82.85 & 82.92\\
        & TransFuse~\cite{zhang2021transfuse} & 14.95 & 10.11 & 11.59 & 11.96 & 11.13 & 11.95 & 80.73 & 72.43 & 82.33 & 72.53 & 81.04 & 77.81\\
        & SSFormer~\cite{wang2022stepwise} & 34.66 & 27.74 & 27.06 & 28.49 & 25.97 & 28.78 & 85.92 & 84.44 & 88.00 & 78.80 & 83.37 & 84.11\\
        & Swin-UNet~\cite{cao2022swin} & 07.90 & 04.27 & 04.29 & 08.08 & 07.32 & 06.37 & 48.16 & 53.91 & 50.00 & 49.57 & 61.68 & 52.66\\
        & SAM~\cite{kirillov2023segment} & 03.81 & 03.66 & 01.47 & 01.40 & 02.89 & 02.64 & 44.65 & 40.57 & 49.94 & 56.43 & 46.31 & 47.58\\
        & SAM (box)~\cite{kirillov2023segment} & 03.99 & 02.78 & 02.60 & 02.67 & 02.55 & 02.92 & 77.21 & 80.40 & 82.65 & 77.73 & 78.60 & 79.32\\
        & SAM (point)~\cite{kirillov2023segment} & 10.49 & 07.48 & 05.16 & 04.66 & 07.79 & 07.12 & 72.53 & 73.27 & 75.30 & 67.80 & 72.08 & 72.20\\
        & MedSAM~\cite{ma2023segment} & 04.07 & 03.29 & 03.16 & 02.86 & 03.11 & 03.30 & 76.46 & 79.26 & 80.97 & 75.26 & 78.70 & 78.13\\
        & \textbf{EFCNet(Ours)} & \textbf{49.10} & \textbf{44.91} & \textbf{43.72} & \textbf{43.46} & \textbf{44.95} & \textbf{45.23} & \textbf{89.11} & \textbf{90.74} & \textbf{89.20} & \textbf{85.39} & \textbf{83.58} & \textbf{87.60}\\
        \hline
        \multirow{14}*{IoU ($\%$)} 
        & U-Net~\cite{ronneberger2015u} & 28.83 & 24.09 & 24.20 & 24.12 & 26.90 & 25.63 & 65.98 & 68.77 & 73.42 & 63.67 & 58.63 & 66.09\\
        & U-Net-Large~\cite{ronneberger2015u} & 29.99 & 25.09 & 25.21 & 25.13 & 27.88 & 26.66 & 68.93 & 69.99 & 76.75 & 62.55 & 65.98 & 68.84\\
        & Attn-UNet~\cite{oktay2018attention} & 30.75 & 27.47 & 26.27 & 25.95 & 26.68 & 27.42 & 72.69 & 72.59 & 73.30 & 68.58 & 68.38 & 71.11\\
        & MSU-Net~\cite{su2021msu} & 30.20 & 27.98 & 25.60 & 25.48 & 25.76 & 27.00 & 79.51 & 78.79 & 77.54 & 72.72 & 68.21 & 75.35\\
        & CaraNet~\cite{lou2022caranet} & 10.36 & 08.04 & 07.21 & 07.77 & 07.50 & 08.18 & 77.24 & 72.04 & 81.40 & 67.29 & 75.56 & 74.71\\
        & TransUNet~\cite{chen2021transunet} & 22.00 & 19.87 & 19.55 & 21.02 & 21.21 & 20.73 & 75.40 & 75.29 & 81.02 & 68.83 & 75.20 & 75.15\\
        & TransFuse~\cite{zhang2021transfuse} & 09.05 & 05.75 & 06.58 & 07.15 & 06.25 & 06.96 & 70.77 & 64.74 & 74.47 & 65.13 & 73.01 & 69.62\\
        & SSFormer~\cite{wang2022stepwise} & 22.65 & 17.67 & 16.57 & 17.95 & 16.02 & 18.17 & 78.16 & 77.30 & 80.93 & 72.74 & 75.52 & 76.93\\
        & Swin-UNet~\cite{cao2022swin} & 05.03 & 02.38 & 02.32 & 05.08 & 04.16 & 03.79 & 37.36 & 41.84 & 39.23 & 38.00 & 50.10 & 41.31\\
        & SAM~\cite{kirillov2023segment} & 02.14 & 02.11 & 00.77 & 00.73 & 01.62 & 01.47 & 39.17 & 35.44 & 44.97 & 51.01 & 41.50 & 42.42\\
        & SAM (box)~\cite{kirillov2023segment} & 02.09 & 01.45 & 01.34 & 01.40 & 01.31 & 01.52 & 67.04 & 70.62 & 72.97 & 67.71 & 69.34 & 69.54\\
        & SAM (point)~\cite{kirillov2023segment} & 06.86 & 04.49 & 03.10 & 02.65 & 04.85 & 04.39 & 64.17 & 65.89 & 67.54 & 60.15 & 65.60 & 64.67\\
        & MedSAM~\cite{ma2023segment} & 02.13 & 01.71 & 01.66 & 01.49 & 01.64 & 01.73 & 66.47 & 69.21 & 71.61 & 65.59 & 69.13 & 68.40\\
        & \textbf{EFCNet(Ours)} & \textbf{35.06} & \textbf{31.45} & \textbf{29.84} & \textbf{29.35} & \textbf{30.25} & \textbf{31.19} & \textbf{82.54} & \textbf{83.71} & \textbf{82.08} & \textbf{76.98} & \textbf{75.59} & \textbf{80.18}\\
        \hline
    \end{tabular}

        \vspace{-0.1in }
\end{table*}

% As shown in Tab.~\ref{big_tab}, our model achieves the best performance in all folds  on both S-HRD and S-Polyp in terms of DSC and IoU, compared to previous SOTA models.

As shown in Tab.~\ref{big_tab}, our model consistently outperforms previous state-of-the-art models across all folds for both S-HRD and S-Polyp, measured by DSC and IoU.

On S-HRD, 
our model demonstrates a noteworthy improvement of
 4.88$\%$ in DSC and 3.77$\%$ in  IoU 
 compared to earlier methods. Similarly, on S-Polyp, our model exhibits a performance boost of 3.49$\%$ in DSC and 3.25$\%$ in IoU. 

 % Tab.~\ref{big_tab} reveals that methods generally struggle with S-HRD, likely due to the smaller size of objects compared to S-Polyp, leading to less available information in the images. Despite these challenges, our model consistently outperforms all others. The CSAA and MPS modules effectively mitigate information loss in the encoder and decoder for small medical objects, enhancing the segmentation capability of our model.

Table~\ref{big_tab} shows that methods tend to perform poorly on S-HRD. That is because the objects in S-HRD are smaller in size compared with S-Polyp, which means that there is less information available in images. Nevertheless, our model still performs best among all methods. 

 Furthermore, the results on S-HRD and S-Polyp indicate that the smaller the medical objects in the datasets, the more significant the improvement of our model compared to previous SOTA methods. This underscores the superiority of our model in small medical object segmentation.

% Moreover, through the experimental results on S-HRD and S-Polyp, it can be found that the smaller the medical objects in datasets, the higher the improvement of our model compared with other methods, which demonstrates the superiority of our model in small medical objects segmentation.

Additionally, simply increasing the size of U-Net (U-Net-Large) yields only marginal improvements in segmentation performance compared to the standard-sized U-Net. In contrast, our EFCNet demonstrates substantial improvement. This indicates that the superior performance of EFCNet in segmentation is primarily attributed to our model design rather than the larger model size. While the addition of CSAA and MPS increases the model's cost, we believe the improvement justifies the associated costs in the realm of small medical object segmentation. We provide detailed model costs comparison in the Supplementary.

% In addition, we compare the cost of the different models in Tab.~\ref{modelcost}. As can be seen, simply increasing the size of U-Net (U-Net-Large) only has a slight improvement in segmentation performance compared to the normal size U-Net, while our EFCNet has a large improvement. This proves that the superior performance of our EFCNet on segmentation is mainly due to our model design rather than the larger model size. Although the addition of CSAA and MPS will increase the cost of the model, we believe that the cost is acceptable, given the significant improvement of our method in the field of small medical object segmentation.

\subsection{Visualization}

\begin{figure*}
    \centering
    \includegraphics[width=17.2cm]{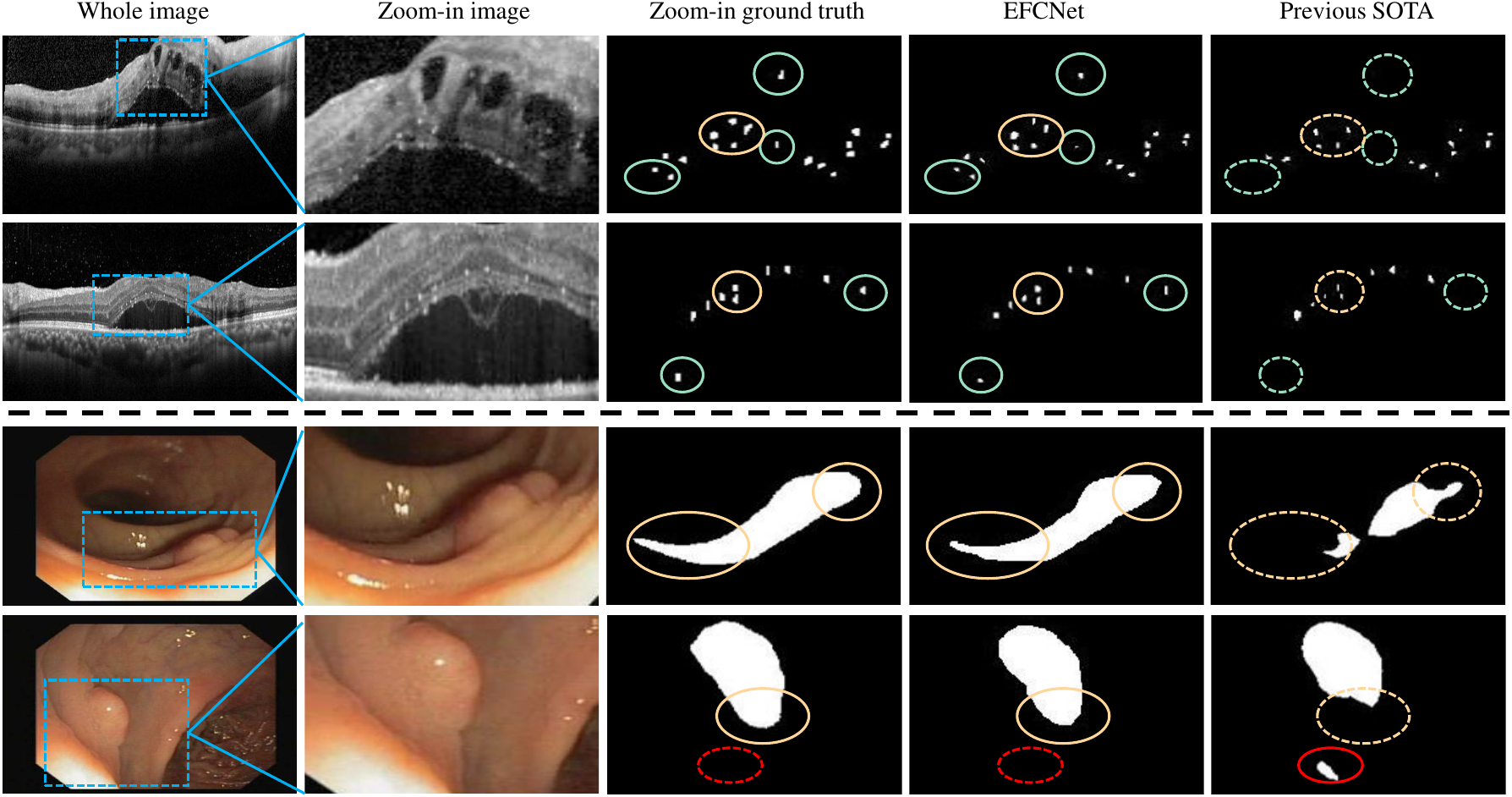}
        \vspace{-0.1in}
    \caption{Visualization of EFCNet (ours) and previous SOTA methods on S-HRD and S-Polyp. The previous SOTA method on S-HRD is Attn-UNet~\cite{oktay2018attention}, and the previous SOTA method on S-Polyp is SSFormer~\cite{wang2022stepwise}. The green circle areas show extremely small medical objects captured by our method that are not captured by previous SOTA methods. The yellow circle areas show that the segmentation of the boundaries of small medical objects in our method is significantly better than the previous SOTA method. The red circle areas show the wrong segmentation of small medical objects in the previous SOTA method while our method is correct.  \label{v_all} }
   
        \vspace{-0.15in}
\end{figure*}

Some visual results of different methods on S-HRD and S-Polyp are shown in Fig.~\ref{v_all}. We compare our EFCNet with the previous SOTA methods. According to our experimental results in Tab.~\ref{big_tab}, the previous SOTA method on S-HRD is Attn-UNet~\cite{oktay2018attention}, and the previous SOTA method on S-Polyp is SSFormer~\cite{wang2022stepwise}. 

% As shown in Fig.~\ref{v_all}, the advantages of our method are mainly reflected in three aspects: (1) Our method is better at capturing extremely small medical objects, as shown in the green circle areas. (2) Our method is more accurate for segmentation in terms of the boundaries of small medical objects, as shown in the yellow circle areas. (3) Our method is much less prone to incorrectly segmenting the background into small medical objects, as shown in the red circle areas.

% On the one hand, CSAA helps our model apply useful local information contained in low-level features of the encoder to the segmentation procedure, ensuring the segmentation ability for fine details of small medical objects. On the other hand, MPS helps our model take advantage of the global perception contained in low-resolution features in the first few stages of the decoder, improving the ability of our model to locate small medical objects.

As illustrated in Fig.~\ref{v_all}, the strengths of our method are predominantly evident in three aspects: (1) Our method excels at capturing extremely small medical objects, as highlighted in the green circle areas.
(2) Our method demonstrates higher accuracy in terms of segmenting the boundaries of small medical objects, as evidenced by the yellow circle areas.
(3) Our method is significantly less prone to erroneously segmenting the background into small medical objects, as indicated by the red circle areas.

On one hand, CSAA facilitates the application of valuable local information from low-level features in the encoder to the segmentation process, ensuring the model's capability to capture fine details of small medical objects. On the other hand, MPS enables the model to leverage global perception inherent in low-resolution features in the initial stages of the decoder, enhancing its ability to locate small medical objects.

\subsection{Ablation Studies}

\begin{table} \small
  \caption{Ablation study on CSAA module and MPS module on S-HRD and S-Polyp in DSC ($\%$) and IoU ($\%$).   \label{abla_all} }
    \vspace{-0.1in}
  \centering
  \tabcolsep=0.03cm
  \begin{tabular}{ccccc}
    \toprule
    \multirow{2}*{Methods} & \multicolumn{2}{c}{S-HRD} & \multicolumn{2}{c}{S-Polyp} \\
    & DSC($\%$) & IoU($\%$) & DSC($\%$) & IoU($\%$)\\
    \midrule
    U-Net & 36.58 & 24.14 & 69.98 & 63.27\\
    U-Net+CSAA & 39.19 & 27.09 & 78.84 & 70.66\\
    U-Net+MPS & 38.98 & 26.85 & 80.20 & 74.14\\
    \textbf{U-Net+CSAA+MPS (Ours)} & \textbf{41.82} & \textbf{29.69} & \textbf{83.26} & \textbf{76.29}\\
    \bottomrule
  \end{tabular}
  \vspace{-0.1in}
\end{table}

We perform several sets of ablation experiments on S-HRD and S-Polyp, confirming the positive effect of CSAA and MPS on segmentation ability for small medical objects respectively.

% We add CSAA and MPS to our U-Net backbone respectively, and the experimental results are shown in the Tab.~\ref{abla_all}. It can be seen that each module helps our model improve its performance. Moreover, after adding both CSAA and MPS, the segmentation ability of our model can be further improved.

\noindent \textbf{Effectiveness of CSAA and MPS}. We incorporate CSAA and MPS into our U-Net backbone individually, and the experimental results are presented in Tab.~\ref{abla_all}. It is evident that each module contributes to the improvement of our model's performance. Furthermore, with the addition of both CSAA and MPS, the segmentation ability of our model experiences further enhancement.

\begin{table} \small
  \caption{Ablation study on the number of stages that CSAA aggregates on S-HRD and S-Polyp in DSC ($\%$) and IoU ($\%$).   \label{abla_csaa} }
    \vspace{-0.1in}
  \centering
  \tabcolsep=0.15cm
  \begin{tabular}{ccccc}
    \toprule
    \multirow{2}*{Methods} & \multicolumn{2}{c}{S-HRD} & \multicolumn{2}{c}{S-Polyp} \\ 
    & DSC ($\%$) & IoU ($\%$) & DSC ($\%$) & IoU ($\%$)\\
    \midrule
    Concat-One & 38.98 & 26.85 & 80.20 & 74.14\\
    AA-One & 40.37 & 28.13 & 82.36 & 75.94\\
    \textbf{AA-All (Ours)} & \textbf{41.82} & \textbf{29.69} & \textbf{83.26} & \textbf{76.29}\\	
    \bottomrule
  \end{tabular}
  \vspace{-0.1in}
\end{table}

\noindent \textbf{Number of Stages in CSAA}.
We change the number of stages aggregated in CSAA module: AA-All aggregates features in all stages of the encoder, which is exactly the CSAA applied in our final model. AA-One only performs axial attention on features in one stage of the encoder. Concat-One only concatenates features in each stage of the encoder to the corresponding decoder without any other processing. The performance of these three methods on S-HRD and S-Polyp is shown in Tab.~\ref{abla_csaa}. It can be seen that among the three models, Concat-One performs the worst. Compared with Concat-One, AA-One can improve the segmentation ability of the model. CSAA aggregates features of all stages of the encoder and performs best among these three methods above.

\begin{table} \small
  \caption{Ablation study on the number of MPS connected to the decoder on S-HRD and S-Polyp in DSC ($\%$) and IoU ($\%$).  \label{abla_mps} }
    \vspace{-0.1in}

  \centering
  \tabcolsep=0.15cm
  \begin{tabular}{ccccc}
    \toprule
    \multirow{2}*{Methods} & \multicolumn{2}{c}{S-HRD} & \multicolumn{2}{c}{S-Polyp} \\
    & DSC ($\%$) & IoU ($\%$) & DSC ($\%$) & IoU ($\%$)\\
    \midrule
    MPS-1 & 39.19 & 27.09 & 78.84 & 70.66\\
    MPS-2 & 40.07 & 27.17 & 82.00 & 73.87\\
    MPS-3 & 40.18 & 28.32 & 82.51 & 75.13\\
    \textbf{MPS-4 (Ours)} & \textbf{41.82} & \textbf{29.69} & \textbf{83.26} & \textbf{76.29}\\
    \bottomrule
  \end{tabular}
  \vspace{-0.1in}
\end{table}

\noindent \textbf{Number of Supervisions in MPS}.
We vary the number of supervisions in the MPS: MPS-4 connects segmentation heads to all stages of the decoder, representing the MPS configuration in our final model.  MPS-3, MPS-2, and MPS-1 connect three, two and one segmentation heads to the decoder respectively. The performance of these four models on S-HRD and S-Polyp is shown in Tab.~\ref{abla_mps}. It is evident that among these four models,  increased supervision correlates with improved model performance.

\section{Conclusion}
We introduce a novel model called EFCNet to address the challenging task of small object segmentation in medical images. EFCNet pays sufficient attention to all features of each stage in the model, effectively reducing the information loss of small medical objects and improving the segmentation accuracy. Specifically, we propose Cross-Stage Axial Attention Module (CSAA) and Multi-Precision Supervision Module (MPS), which alleviate the loss of information in the encoder and decoder respectively,  leading to a substantial enhancement in model performance. Moreover, we establish a new benchmark for small medical object segmentation research. 
Our experiments on two datasets demonstrate that CSAA and MPS contribute to improved segmentation accuracy, with our model significantly outperforming previous state-of-the-art models.

% Our experiment on two datasets demonstrates that CSAA and MPS can help the model improve the accuracy of segmentation, and that our model significantly outperforms previous state-of-the-art models.

{
    \small
    \bibliographystyle{ieeenat_fullname}
    \bibliography{main}
}

% WARNING: do not forget to delete the supplementary pages from your submission 
\clearpage
\setcounter{page}{1}
\maketitlesupplementary

\section{Rationale}
\label{sec:rationale}
Having the supplementary compiled together with the main paper means that:
\begin{itemize}
\item The supplementary can back-reference sections of the main paper, for example, we can refer to \cref{sec:intro};
\item The main paper can forward reference sub-sections within the supplementary explicitly (e.g. referring to a particular experiment); 
\item When submitted to arXiv, the supplementary will already included at the end of the paper.
\end{itemize}
To split the supplementary pages from the main paper, you can use \href{https://support.apple.com/en-ca/guide/preview/prvw11793/mac#:~:text=Delete%20a%20page%20from%20a,or%20choose%20Edit%20%3E%20Delete).}{Preview (on macOS)}, \href{https://www.adobe.com/acrobat/how-to/delete-pages-from-pdf.html#:~:text=Choose%20%E2%80%9CTools%E2%80%9D%20%3E%20%E2%80%9COrganize,or%20pages%20from%20the%20file.}{Adobe Acrobat} (on all OSs), as well as \href{https://superuser.com/questions/517986/is-it-possible-to-delete-some-pages-of-a-pdf-document}{command line tools}.

\section{Overview}
In the supplementary material, we firstly provide additional analysis of model cost in Sec.~\ref{cost}. Then, we provide additional details about the architecture of our EFCNet in Sec.~\ref{model} and experiment settings in Sec.~\ref{epx}. In the end, we introduce ethical considerations in Sec.~\ref{ethic}.

\section{Analysis of Model Cost }
\label{cost}
We provide model cost comparison of our EFCNet and other UNet-based methods in Tab.~\ref{modelcost_sup} and performance improvement compared to U-Net~\cite{ronneberger2015u} in Tab.~\ref{improve_sup}. We can draw a conclusion that in the field of small medical object segmentation, simply increasing the model size like U-Net-Large cannot bring significant improvement in segmentation performance based on the standard-sized U-Net. In comparison, our EFCNet achieves far better segmentation performance than other UNet-based methods with model cost less than that of U-Net-Large. Indeed, our model remains relatively large to effectively address the intricate challenge of the segmentation task.

\section{Additional Details about Model Architecture}
\label{model}
We provide additional details about the backbone network, the CSAA Module and the MPS Module in our EFCNet as shown in Tab.~\ref{backbone_sup}, Tab.~\ref{csaa_sup} and Tab.~\ref{mps_sup} respectively.

\section{Additional Details about Experiment Settings}
\label{epx}

\noindent \textbf{Experimental environment}. The environment of our experiment is as follows. GPU: NVIDIA RTX A6000; CUDA Version: 11.7; Python Version: 3.10.4; Torch Version: 1.13.1.

\noindent \textbf{Split of Datasets}. We perform five-fold cross-validation in our experiments. In each split, datasets are separated into training set, validation set, testing set by a ratio of 7:1:2.

\begin{table} \small
  \caption{Model cost comparison of our EFCNet with other UNet-based methods.   \label{modelcost_sup}}
    \vspace{-0.1in}

  \centering
  \tabcolsep=0.23cm
  \begin{tabular}{ccc}
    \toprule
    Methods & FLOPs (G) & Params (M)\\
    \midrule
    U-Net~\cite{ronneberger2015u} & 91.94 & 37.66\\
    Attn-UNet~\cite{oktay2018attention} & 125.94 & 34.88\\
    MSU-Net~\cite{su2021msu} & 143.05 & 47.09\\
    U-Net-Large~\cite{ronneberger2015u}& 430.57 & 100.37\\
    \midrule
    U-Net+CSAA & 375.52 & 87.37\\
    U-Net+MPS & 101.48 & 38.81\\
    U-Net+CSAA+MPS (EFCNet) & 385.06 & 88.52\\
    \bottomrule
  \end{tabular}

\end{table}

\begin{table*} \small
  \caption{Performance improvement among our EFCNet and other UNet-based methods compared to U-Net~\cite{ronneberger2015u}.   \label{improve_sup}}
  \vspace{-0.05in}

  \centering
  \tabcolsep=0.5cm
  \begin{tabular}{ccccc}
    \toprule
    \multirow{2}*{Methods} & \multicolumn{2}{c}{S-HRD} & \multicolumn{2}{c}{S-Polyp} \\
    & $\Delta$DSC($\%$) & $\Delta$IoU($\%$) & $\Delta$DSC($\%$) & $\Delta$IoU($\%$)\\
    \midrule
    U-Net~\cite{ronneberger2015u} & +0.00&+0.00&+0.00&+0.00\\
    Attn-UNet~\cite{oktay2018attention} & +1.64 & +1.79 & +2.37 & +5.02\\
    MSU-Net~\cite{su2021msu} & +1.62 & +1.37 & +7.34 & +9.26\\
    U-Net-Large~\cite{ronneberger2015u}& +0.77 & +1.03 & +1.89 & +2.75 \\
    U-Net+CSAA+MPS (EFCNet) & \textbf{+6.52} & \textbf{+5.56} & \textbf{+11.18} & \textbf{+14.09} \\
    \bottomrule
    
  \end{tabular}

\end{table*}

\begin{table*} \small
  \caption{Details of the backbone network of our EFCNet.   \label{backbone_sup}}
  \vspace{-0.05in}
  \centering
  \tabcolsep=0.8cm
  \begin{tabular}{ccc}
    \toprule
    Stage & Layer & Output shape\\
    \midrule
    Input & - & (3, 352, 352)\\
    \midrule
    Encoder stage1 & (Conv, BN, ReLU) \(\times\) 2, Downsample & (64, 176, 176)\\
    Encoder stage2 & (Conv, BN, ReLU) \(\times\) 2, Downsample & (128, 88, 88)\\
    Encoder stage3 & (Conv, BN, ReLU) \(\times\) 2, Downsample & (256, 44, 44)\\
    Encoder stage4 & (Conv, BN, ReLU) \(\times\) 2, Downsample & (512, 22, 22)\\
    \midrule
    Decoder stage4 & (Conv, BN, ReLU) \(\times\) 2, Upsample, Concat & (1024, 44, 44)\\
    Decoder stage3 & (Conv, BN, ReLU) \(\times\) 2, UPsample, Concat & (512, 88, 88)\\
    Decoder stage2 & (Conv, BN, ReLU) \(\times\) 2, UPsample, Concat & (256, 176, 176)\\
    Decoder stage1 & (Conv, BN, ReLU) \(\times\) 2, UPsample, Concat & (128, 352, 352)\\
    \bottomrule
  \end{tabular}
  
\end{table*}

\begin{table*} \small
  \caption{Details of the CSAA Module in our EFCNet.   \label{csaa_sup}}
  \vspace{-0.05in}
  \centering
  \tabcolsep=0.8cm
  \begin{tabular}{ccc}
    \toprule
    Stage & Layer & Output shape\\
    \midrule
    \multirow{4}*{CSAA stage1} & Resize: ((Conv, BN, ReLU) \(\times\) 2) & (\(C^{*}\), \(H^{*}\), \(W^{*}\))\\
    & W-CSAA: one-dimensional attention module & (\(C^{*}\), \(H^{*}\), \(W^{*}\))\\
    & H-CSAA: one-dimensional attention module & (\(C^{*}\), \(H^{*}\), \(W^{*}\))\\ 
    & Resize back: ((Conv, BN, ReLU) \(\times\) 2) & (64, 352, 352)\\
    \midrule
    \multirow{4}*{CSAA stage2} & Resize: ((Conv, BN, ReLU) \(\times\) 2) & (\(C^{*}\), \(H^{*}\), \(W^{*}\))\\
    & W-CSAA: one-dimensional attention module & (\(C^{*}\), \(H^{*}\), \(W^{*}\))\\
    & H-CSAA: one-dimensional attention module & (\(C^{*}\), \(H^{*}\), \(W^{*}\))\\ 
    & Resize back: ((Conv, BN, ReLU) \(\times\) 2) & (128, 176, 176)\\
    \midrule
    \multirow{4}*{CSAA stage3} & Resize: ((Conv, BN, ReLU) \(\times\) 2) & (\(C^{*}\), \(H^{*}\), \(W^{*}\))\\
    & W-CSAA: one-dimensional attention module & (\(C^{*}\), \(H^{*}\), \(W^{*}\))\\
    & H-CSAA: one-dimensional attention module & (\(C^{*}\), \(H^{*}\), \(W^{*}\))\\ 
    & Resize back: ((Conv, BN, ReLU) \(\times\) 2) & (256, 88, 88)\\
    \midrule
    \multirow{4}*{CSAA stage4} & Resize: ((Conv, BN, ReLU) \(\times\) 2) & (\(C^{*}\), \(H^{*}\), \(W^{*}\))\\
    & W-CSAA: one-dimensional attention module & (\(C^{*}\), \(H^{*}\), \(W^{*}\))\\
    & H-CSAA: one-dimensional attention module & (\(C^{*}\), \(H^{*}\), \(W^{*}\))\\ 
    & Resize back: ((Conv, BN, ReLU) \(\times\) 2) & (512, 44, 44)\\
    \bottomrule
  \end{tabular}
  
\end{table*}

\begin{table*} \small
  \caption{Details of the MPS Module in our EFCNet.   \label{mps_sup}}
  \vspace{-0.05in}
  \centering
  \tabcolsep=0.6cm
  \begin{tabular}{ccc}
    \toprule
    Stage & Layer & Output shape\\
    \midrule
    MPS stage1 & Segmentation: (Conv, BN, ReLU) \(\times\) 2, Conv, Sigmoid & (1, 352, 352)\\
    \midrule
    \multirow{2}*{MPS stage2} & Segmentation: (Conv, BN, ReLU) \(\times\) 2, Conv, Sigmoid & (1, 176, 176)\\
    & Upsample: Nearest interpolation & (1, 352, 352)\\
    \midrule
    \multirow{2}*{MPS stage3} & Segmentation: (Conv, BN, ReLU) \(\times\) 2, Conv, Sigmoid & (1, 88, 88)\\
    & Upsample: Nearest interpolation & (1, 352, 352)\\
    \midrule
    \multirow{2}*{MPS stage4} & Segmentation: (Conv, BN, ReLU) \(\times\) 2, Conv, Sigmoid & (1, 44, 44)\\
    & Upsample: Nearest interpolation & (1, 352, 352)\\
    \bottomrule
  \end{tabular}
  
\end{table*}

\section{Ethical Considerations}
\label{ethic}

The collection and utilization of human data in our research project adheres to the highest ethical standards. Our study has received the full approval of the Institutional Review Board and the Ethics Committee of a hospital. This approval process is conducted in strict accordance with the principles outlined in the Declaration of Helsinki~\cite{williams2008declaration}, which provides ethical guidelines for medical research involving human subjects. All recruited patients have signed the informed consent to publish this paper. Here, we provide the information on data collection and annotation, appropriate consent and privacy considerations. 

\noindent \textbf{Data collection and annotation}. In this investigation, retinal OCT scans are collected from eyes of 313 patients who seek treatment for macular edema associated with diabetic retinopathy or retinal vein occlusion at a hospital within the past six months. The investigation has been approved by the Institutional Review Board and the Ethics Committee of a hospital, in accordance with the principles of the Declaration of Helsinki~\cite{williams2008declaration}. Each OCT scan is centered on the fovea, either vertically or horizontally. Macular edema is defined as a central retinal thickness (CRT) greater than 300 mm. Small hyperreflective dots (small HRDs) are defined as discrete tiny dots with diameter between 20 micron and 40 micron, characterized by reflectivity similar to that of the nerve fiber layer and the absence of back shadowing~\cite{huang2021algorithm}. The ground truths of small HRDs have been manually labeled by experienced eye doctors with over ten years of expertise. 

\noindent \textbf{Appropriate consent and privacy considerations}. We confirm that appropriate consent has been obtained for the use and display of images in our research. To uphold the privacy and confidentiality of the individuals involved, we take rigorous measures to ensure that all identifying information, including names, genders, and birth dates of patients, has been thoroughly removed from the images prior to any processing or analysis. These steps are taken to safeguard patient privacy and adhere to ethical standards. We understand the critical importance of addressing privacy concerns when dealing with medical images, and we are committed to upholding the highest ethical standards in our research practices.

\medskip

\end{document}